\documentclass{appolb}
\usepackage{graphicx}
\usepackage{cite}
\newcommand{\wasa}{\mbox{WASA-at-COSY}}
\begin{document}
\title{Search for $\eta$-mesic nuclei with \wasa\footnote{Presented at the Zakopane Conference on Nuclear Physics ''Extremes of the Nuclear Landscape``, Zakopane, Poland, August 31--September 7, 2014.} 
}
\author{Wojciech Krzemie\'n\footnote{Present address: High Energy Physics Division, National Centre for Nuclear Research, 05-400 Otwock-\'Swierk, Poland}, P. Moskal and M. Skurzok for the \wasa~collaboration
\address{M. Smoluchowski Institute of Physics, Jagiellonian University, 30-059 Cracow, Poland}
}
\maketitle
\begin{abstract}
We search for an evidence of $\eta$-mesic He with the WASA detector. 
Two dedicated experiments were performed at the 
Cooler Synchrotron COSY-J\"ulich. 
The experimental method is based on the measurement of the excitation functions 
for the two reaction channels: $dd \rightarrow {^3\mbox{He}} p \pi^{-}$ 
and $dd \rightarrow {^3\mbox{He}} n \pi^{0}$, where the outgoing N-$\pi$ pairs 
originate from the conversion of the $\eta$ meson on a nucleon inside the 
He nucleus. 
In this contribution, the experimental method is shortly described and  
the current status of the analysis is presented.
\end{abstract}
  
\section{Introduction}
One can define exotic atoms and nuclei, as systems in which 
one of the standard component particles is replaced by 
an exotic particle e.g. pionic atoms, where the negatively charged pion replaces 
an electron.
The studies of exotic systems were proved to be very fruitful in the past, 
e.g. experiments on hypernuclei started a new branch of investigations - 
the strangess physics. 
More recently, studies of meson-nucleus interaction   
have attracted a lot of interest 
because they serves not only to better understand the 
meson - nucleon interaction but also provide 
information about meson properties embedded in nuclear matter, 
which are directly linked with the postulated partial 
restoration of the chiral symmetry and the structure of the QCD vacuum~\cite{hiren,BassTom,Jido,InoueOset,metag,tshushima1,tshushima2, Friedman_2013, Itahashi,eta-prime-mesic-Nagahiro, eta-prime-mesic-Nagahiro-Oset, ELSA-MAMI-plan-Krusche, Krusche2, Machner, moskalsymposium}.
In general meson - nucleon binding is the result of an interplay 
of electromagnetic and strong forces but in the case of neutral mesons it is exclusively 
due to the strong interaction, thus the mesic nucleus
can be considered as a meson captured in the mean field of the nucleons.

The $\eta$-mesic nucleus is one of the most promising candidates for such a state 
because of the relatively strong $\eta$-nucleon interaction~\cite{wycech,moskal}.
Already in 1986, 
Haider and Liu postulated the hypothesis of a $\eta$-mesic nucleus~\cite{HaiderLiu1}.
Since then many tries have been undertaken to experimentally confirm 
its existence but without any conclusive result.

\section{Experimental method}
The search for the $\eta$-mesic bound states can be divided into 
two categories. In-direct search methods consist of  
study $\eta$ production cross-section right above the threshold 
to infer its subthreshold behaviour, and  to establish binding conditions 
e.g. expressed as $\eta$-nucleus scattering length. 
Although, such studies~\cite{jurek-he3,timo,mami2, gem} provided important experimental indications in the case of 
${^3\mbox{He}}$ and ${^4\mbox{He}}$ systems, where they showed the existence 
of a s-wave pole in the scattering matrix, 
they are not able to give the decisive answer whether the pole corresponds to a virtual 
or bound state as it was stated by~\cite{Wilkin_Acta2014}.
The second category contains direct-search methods, which look for a subthreshold manifestation of a bound state in the excitation
functions for chosen decay modes. 
This approach impose special experimental requirements e.g.
very accurate knowledge of the total reaction energy, 
and good control over the luminosity and acceptance for consecutive energy bins.

Both aforementioned conditions are fulfilled with the WASA detector at COSY synchrotron.
The WASA detection system~\cite{wasa} provide a high acceptance 
combined with the possiblity of registering all final state particles.
Also, we take advantage of the COSY synchrotoron ramped beam mode, 
which permits to smoothly change the beam momentum within one acceleration cycle, 
and in consequence to  obtain an excellent reaction energy resolution.
We carry out the search of a $\eta$-mesic helium produced in proton-deuteron and  deuteron-deuteron collisions.
We concentrate on the $\eta-{\mbox{He}}$ decay mode,  in which 
the trapped $\eta$ meson is absorbed on one of the 
nucleons in the He nucleus. The nucleon is excited to 
the  $N^{*}$ (1535) state, which subsequently decays into a pion-nucleon pair.
In case of the $dd \rightarrow (\eta-{^{4}\mbox{He}})_{bound}$ channel, the remaining three nucleons are likely to form a $^3$He
or $^3$H nucleus. The outgoing $^3\mbox{He}$ nucleus is expected to have a  
relatively low momentum in the center of mass (c.m.) frame that can be approximated 
by the Fermi momentum distribution of the nucleons inside the $^4\mbox{He}$ nucleus.
The process described above should result in a resonance-like structure in the excitation function of the $dd \rightarrow ^3$He$ p \pi^{-}$ and the $dd \rightarrow ^3$He$ n \pi^{0}$ reactions if we select events with low ${^3}$He center-of-mass (c.m.) momenta.

\section{Experiments}
So far, three dedicated measurements were done with \wasa.
The first experiment was performed in June 2008 by measuring the excitation fun\-ction of the $dd \rightarrow$ $^{3}\hspace{-0.03cm}\mbox{He} p \pi^{-}$  reaction near the $\eta$ meson production threshold. 
An upper limit for the formation and decay of the bound state in the process
$dd \rightarrow ({{^4\mbox{He}}-\eta})_{bound} \rightarrow {^3\mbox{He}} p \pi^{-}$ 
at the 90\% confidence level, was determined from 20~nb to 27~nb for the bound state width ranging from 5~MeV to 35~MeV, respectively ~\cite{Adlarson2013}.
During the second experiment, in November 2010, two channels of the $\eta$-mesic helium decay were registered :  $dd\rightarrow(^{4}\mbox{He}$-$\eta)_{bound}\rightarrow$ $^{3}\mbox{He} p \pi{}^{-}$ and  $dd\rightarrow(^{4}\mbox{He}$-$\eta)_{bound}\rightarrow$ $^{3}\mbox{He} n \pi{}^{0} \rightarrow$ 
${^{3}\mbox{He}} n \gamma \gamma$~\cite{Magda2, WKrzemien_2014, Moskal_FewBody} in the excess energy range from -70~MeV to 30~MeV.

\begin{figure}[htb]
\centering
\includegraphics[width=5.5cm,height=5cm]{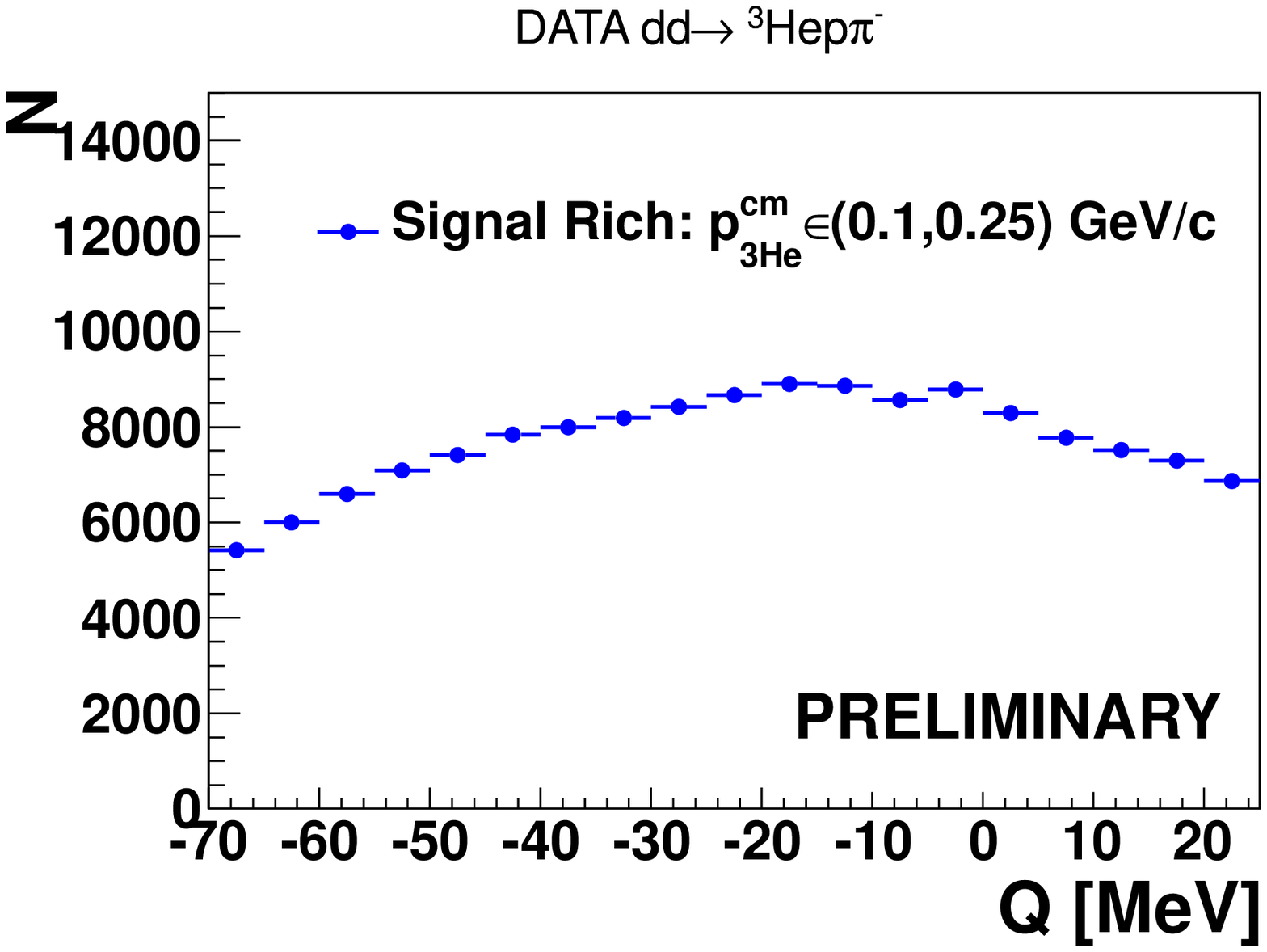}
\includegraphics[width=5.5cm,height=5cm]{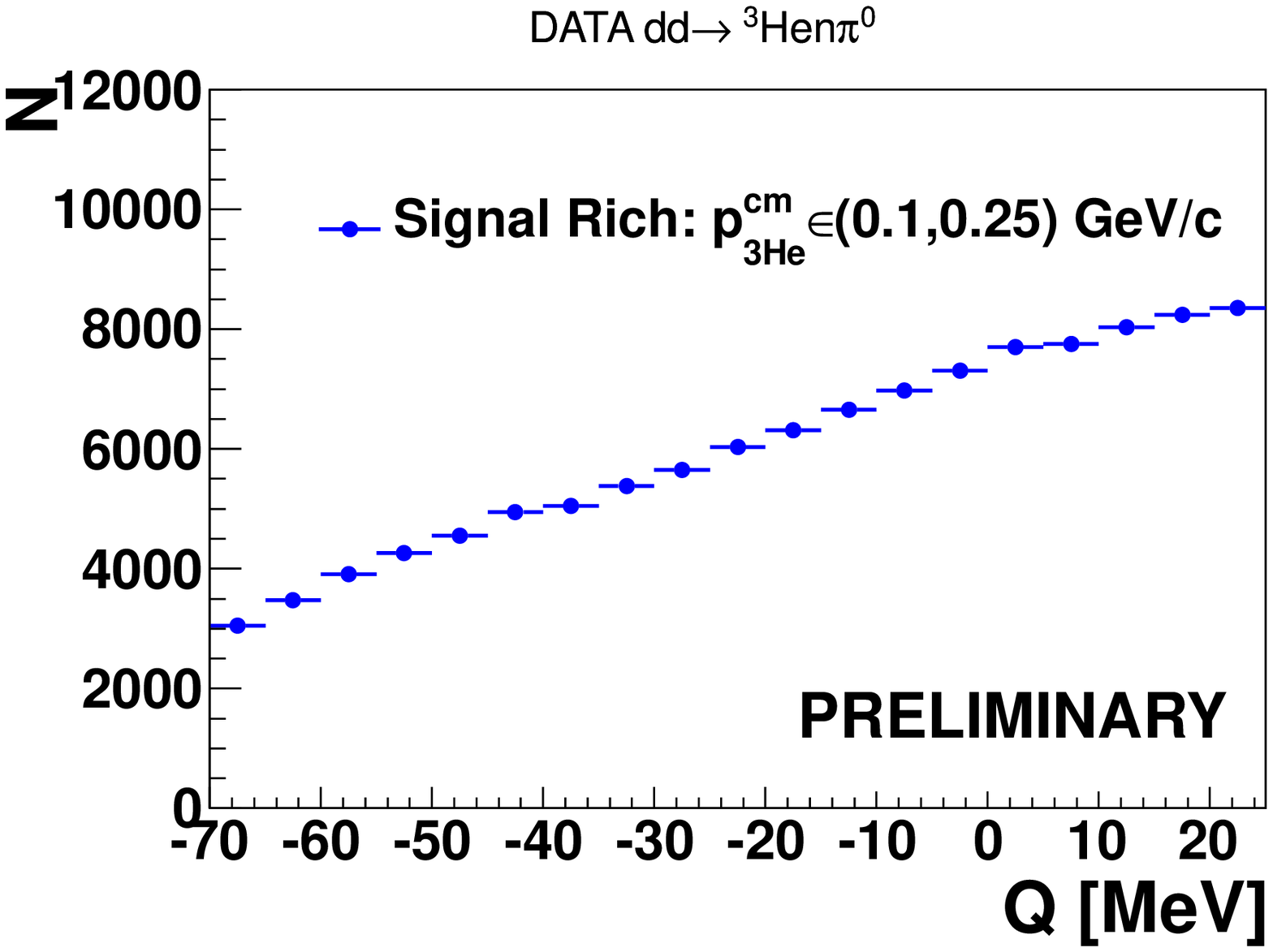}
\caption{\label{fig:excit2}Preliminary excitation function for the $dd \rightarrow {^3\mbox{He}} p \pi^-$  and  for the $dd \rightarrow {^3\mbox{He}} n \pi^{0}$ reactions under condition that the ${^3\mbox{He}}$ momentum in c.m. frame is in the range from 0.1 to 0.25 GeV/c ("signal-rich" area). The distributions are not corrected for efficiency. }

\end{figure}

The preliminary excitation functions for the "signal-rich" region, corresponding to the low ${^3\mbox{He}}$ momenta c.m. frame,  in which we expect the highest signal to noise ratio, are presented in Fig.~\ref{fig:excit2}.

The  predictions given in~\cite{Wycech-Acta}, state a cross-section of 4.5 nb. This can be confronted 
with the expected sensitivity from the 2010 data, which is of the order of few nb. 
Therefore, the ongoing analysis should be able to reveal the hypothetical signal 
from the decay of mesic nucleus in ${^{4}\mbox{He}}$ state.

\section{Future prospects}
In May 2014 we carried out the third complementary experiment 
in proton deuteron collisions, aiming at the exploration of the
${^{3}\mbox{He}}$ mesic nuclei. 
This was motivated by the recent experimental and theoretical results
~\cite{Friedman_2013, Wilkin_Acta2014, Gal_2014, Mesic-Kelkar, Krusche_2014}, 
which favor $\eta-{^{3}\mbox{He}}$ over $\eta-{^{4}\mbox{He}}$ bound states.
In addition to the previously described decay mode via N$^{\star}$ resonance, 
we also considered a second mechanism, in which the bound $\eta$  
decays, while still "orbiting" around a nucleus e.g. via
$pd \rightarrow$ (${^{3}\mbox{He}}$-$\eta)_{bound} \rightarrow {^{3}\mbox{He}} 6\gamma$ 
reaction.
Although, the predicted cross-section for this decay mode is relatively small 
(0.4 nb ~\cite{Wilkin_Acta2014}) the background is expected to  be highly supressed.
A week-long measurement  with an average luminosity of about 6$\cdot10^{30}$ cm$^{-2}$ s$^{-1}$ allowed to collect high statistics of data. The analysis is in progress.

\section{Acknowledgments}
This work was supported by the Foundation for Polish Science - MPD programme, 
co-financed by the European Union within the European Regional Development Fund, 
by the Polish National Science Center under grant No. 2011/01/B/ST2/00431 and 
by the FFE grants of the Research Center J\"ulich.

%%%%%%%%%%%%%%%%%%%%%%%%%%%%%%%%%%%%%

%uncomment the following lines to place a figure
%\begin{figure}[htb]
%\centerline{%
%\includegraphics[width=12.5cm]{Fig1}}
%\caption{Plot of ...}
%\label{Fig:F2H}
%\end{figure}

\end{document}